\documentclass[sigconf,nonacm]{acmart}

\setcopyright{none}
\renewcommand\footnotetextcopyrightpermission[1]{}
\usepackage{algorithm}
\usepackage{algpseudocode}
\usepackage{listings}
\usepackage{xcolor}
\usepackage{booktabs}
\usepackage{multirow}
\usepackage{subcaption}
\usepackage{amsmath}

\usepackage{amssymb}
\usepackage{tikz}
\usetikzlibrary{arrows.meta, positioning}

\definecolor{cobolbg}{RGB}{245,245,240}
\definecolor{pythonbg}{RGB}{240,245,250}
\definecolor{z3bg}{RGB}{250,240,245}
\definecolor{commentgreen}{RGB}{0,128,0}
\definecolor{keywordblue}{RGB}{0,0,180}

\lstdefinestyle{cobolstyle}{
  backgroundcolor=\color{cobolbg},
  basicstyle=\ttfamily\scriptsize,
  keywordstyle=\bfseries\color{keywordblue},
  commentstyle=\color{commentgreen},
  numbers=left,
  numberstyle=\tiny\color{gray},
  numbersep=5pt,
  frame=single,
  breaklines=true,
  captionpos=b,
  morekeywords={IDENTIFICATION,DIVISION,PROGRAM-ID,DATA,WORKING-STORAGE,
    SECTION,PROCEDURE,EVALUATE,TRUE,WHEN,OTHER,END-EVALUATE,COMPUTE,
    STOP,RUN,PIC,VALUE,ZEROS,IF,ELSE,END-IF,PERFORM,MOVE,TO,ADD,
    SUBTRACT,MULTIPLY,DIVIDE,CALL,EXEC,SQL,END-EXEC,OPEN,READ,WRITE,
    CLOSE,GO,COPY}
}

\lstdefinestyle{pythonstyle}{
  backgroundcolor=\color{pythonbg},
  basicstyle=\ttfamily\scriptsize,
  keywordstyle=\bfseries\color{keywordblue},
  commentstyle=\color{commentgreen},
  stringstyle=\color{red!70!black},
  numbers=left,
  numberstyle=\tiny\color{gray},
  numbersep=5pt,
  frame=single,
  breaklines=true,
  captionpos=b,
  language=Python
}

\lstdefinestyle{z3style}{
  backgroundcolor=\color{z3bg},
  basicstyle=\ttfamily\scriptsize,
  numbers=left,
  numberstyle=\tiny\color{gray},
  numbersep=5pt,
  frame=single,
  breaklines=true,
  captionpos=b
}

\begin{document}

\title{IRONPROOF: COBOL-to-Python Transpilation with SMT-Based Equivalence Checking}

\author{Dominik Blain}
\affiliation{%
  \institution{Ironproof}
  \country{Canada}
}
\email{dominik@ironproof.ai}


\begin{abstract}
COBOL remains in production in financial institutions and government agencies, and translating it to a modern language risks changing what the programs compute. Large language models (LLMs) can translate code, but their output carries no guarantee of semantic equivalence, and a divergence on a few inputs can pass conventional testing.

We present \textsc{Ironproof}, an end-to-end pipeline that combines rule-based COBOL-to-Python transpilation with SMT-based formal equivalence verification using the Z3 theorem prover. Given a COBOL program, \textsc{Ironproof} parses the source into an intermediate representation covering common enterprise COBOL constructs, generates Python with \texttt{Decimal}-typed arithmetic and \texttt{@dataclass} working storage, then encodes both programs as Z3 formulas over shared symbolic inputs and checks whether any input witnesses a semantic divergence. If no divergence exists (UNSAT), \textsc{Ironproof} emits a machine-checkable equivalence certificate; if Z3 produces a counterexample (SAT), the tool can return it to an LLM in a correction loop, which this paper describes but does not evaluate.

We evaluate \textsc{Ironproof} on a benchmark of 2{,}345 COBOL programs drawn from the GnuCOBOL compiler test suite (1{,}632 programs), the NIST CCVS85 validation suite, open-source enterprise collections, and systematically generated construct variations. The parser accepts all 2{,}345 programs; our Python generator refuses 400 rather than emit Python it cannot stand behind. A program enters the checking path when its COBOL side yields at least one Z3-encodable output, and 782 do. Of these, \textbf{606 (77.5\%)} are proved equivalent, 101 are only partially verified, none is refuted, and 75 fail inside our own pipeline; the remaining 1{,}563 yield no encodable output (file I/O, embedded SQL, PERFORM loops, INSPECT) and are reported separately. Our failures stay in the denominator, since a rate that excludes them cannot fall. Two qualifications bound what these proofs mean. The first is provenance: on independently authored programs (GnuCOBOL, NIST, dscobol, community---not written by us) the rate is \textbf{52.6\%} (153 of 291), against 92.3\% on programs we wrote. The second is the input domain. \textbf{37 of the 153 independent proofs} are on programs that read input---a file, \texttt{ACCEPT}, parameters---that the encoder does not model, so each covers a single execution of a program that has many. Only \textbf{14} proofs quantify over an input that a proved output depends on, and 15 more have free inputs that no proved output depends on. The remaining 540 are on closed programs, which have one execution and are fully covered by it. On two public business-application corpora (AWS CardDemo and IBM GenApp, 75 programs), the encoder models no program end to end and 1 of 712 paragraphs. A baseline run in April 2026 against LLM-only translation found that the LLM failed on 57\% of 94 independently authored programs, counting both Z3-confirmed semantic errors and translations our encoder could not verify. We also report PIC-bounded overflow detection, which flags truncation risks the unbounded encoding does not model. Finally, an earlier confrontation of the proofs with GnuCOBOL 3.2.0, on the pipeline of 2026-08-26, found that \textbf{24 of the 49} independently authored proved programs it could execute disagreed with the runtime on a variable inside the proof's own domain. These figures delimit what an emission-fidelity proof establishes.
\end{abstract}

\ccsdesc[500]{Software and its engineering~Software verification and validation}
\ccsdesc[300]{Software and its engineering~Compilers}

\keywords{COBOL modernization, LLM code translation, formal verification, SMT solving, translation validation, program equivalence}

\maketitle


\section{Introduction}\label{sec:intro}

COBOL, designed in 1959, still runs core systems in banking and government; Reuters reported in 2017 that \$3 trillion in daily commerce flows through COBOL systems~\cite{reuters2017cobol}.

Modernization translates such code to a language like Java or Python. It requires specialists who understand both ecosystems~\cite{sneed2010automated}, and it can introduce semantic regressions: a misread boundary condition, a rounding difference, or a reordered branch changes results without failing a test that does not exercise it.

Large language models (LLMs) are used for code translation: TransCoder~\cite{roziere2020transcoder} translates between programming languages, and code models such as Codex~\cite{chen2021evaluating} and StarCoder~\cite{li2023starcoder} generate code from natural language. Commercial offerings including IBM watsonx Code Assistant for Z~\cite{ibm2024watsonx} and AWS Mainframe Modernization~\cite{aws2023mainframe} specifically target COBOL migration. LLMs provide no formal guarantee about the semantic correctness of a translation. Pan et al.\ find that LLM translations between C, C++, Go, Java and Python frequently introduce bugs~\cite{pan2024cobol}; such errors can produce correct outputs on most inputs and diverge on specific ones. In financial and governmental code, such an error changes computed amounts for every input it affects.

Formal methods offer the complementary guarantee that LLMs lack. Translation validation, as pioneered by Pnueli et al.~\cite{pnueli1998translation} and implemented in systems such as CompCert~\cite{leroy2009compcert}, Alive2~\cite{lopes2021alive2}, and CertiCoq~\cite{anand2017certicoq}, provides machine-checkable proofs that a transformed program preserves the semantics of its source. However, applying formal verification to legacy COBOL migration faces its own challenges: the semantic gap between COBOL and modern languages is large, COBOL's data model (PIC clauses, group items, REDEFINES) has no direct analog in most target languages, and existing verification infrastructure does not natively support COBOL.

\textsc{Ironproof} combines rule-based transpilation with SMT-based equivalence checking and LLM-driven error correction: the rule-based generator produces the translation, the solver either proves it equivalent to our representation of the COBOL or returns a counterexample, and an LLM can be asked to repair a translation the solver refutes.

\paragraph{Industrial context.}
Industrial modernization offerings from IBM~\cite{ibm2024watsonx} and AWS~\cite{aws2023mainframe} emphasize automation and AI-assisted code transformation, and industry guidance names failure risk, cost overruns, operational disruption, regulation and talent shortages as constraints~\cite{deloitte2024banking}. \textsc{Ironproof} does not replace migration tooling. It addresses one question within it: whether translated business logic is semantically equivalent to the original COBOL.

\smallskip\noindent\textbf{Contributions.} This paper makes three contributions:

\begin{enumerate}
  \item \textbf{A transpilation pipeline with SMT-based equivalence checking.} We present \textsc{Ironproof}, a system that parses COBOL into an intermediate representation, generates Python via rule-based transpilation, encodes both programs as Z3 formulas over shared symbolic inputs, and checks semantic equivalence. The tool also includes a loop that returns Z3 counterexamples to an LLM for correction; it is described but not evaluated here (\S\ref{sec:approach}).

  \item \textbf{An SMT encoding of the COBOL computational core and its Python translation.} We define a parser and Python code generator covering 10 COBOL construct types (arithmetic, conditionals, EVALUATE, PERFORM loops, MOVE, CALL, EXEC SQL, file I/O, GO TO, and COPY books). Of these, four form the Z3-encodable fragment over which we establish formal equivalence---arithmetic operations (COMPUTE, ADD/SUBTRACT/MULTIPLY/DIVIDE), MOVE, IF/ELSE, and EVALUATE---together with bounded compositions of them. The remaining six (PERFORM loops, CALL, EXEC SQL, file I/O, GO TO, and COPY books) are detected and not encoded; an output that depends on them is left unverified rather than proved (\S\ref{sec:z3encoding}). Some proved programs still contain such constructs, flattened identically on both sides by the shared parser; \S\ref{sec:noncircular} counts them (31 of 606).

  \item \textbf{A large-scale empirical evaluation with LLM baseline.} We evaluate \textsc{Ironproof} on 2{,}345 COBOL programs, including 1{,}632 from the GnuCOBOL compiler test suite and 69 from the NIST CCVS85 validation suite. We prove equivalence on 606 of the 782 programs that enter the checking path (77.5\%; 52.6\% on independently authored code), state how many of those proofs cover a single execution of a program that reads input (37) and how many quantify over an input that a proved output depends on (14), and compare against an LLM-only baseline run in April 2026, which fails on 54 of 94 independently authored programs (57\%), counting translations our encoder could not verify as failures. We also introduce PIC-bounded overflow detection (\S\ref{sec:evaluation}).
\end{enumerate}


\section{Motivating Example}\label{sec:motivating}

Consider a simplified tax-bracket calculation. The COBOL source uses an \texttt{EVALUATE TRUE} construct to select the applicable marginal tax rate based on income brackets:

\begin{lstlisting}[style=cobolstyle,caption={COBOL tax bracket calculation using EVALUATE TRUE with $\leq$ boundaries.},label={lst:cobol-tax}]
       IDENTIFICATION DIVISION.
       PROGRAM-ID. FED-TAX-CALC.
       DATA DIVISION.
       WORKING-STORAGE SECTION.
       01 INCOME   PIC 9(8)V99.
       01 TAX-RATE PIC V9(4).
       01 TAX-AMT  PIC 9(8)V99.
       PROCEDURE DIVISION.
           EVALUATE TRUE
               WHEN INCOME <= 47150
                   COMPUTE TAX-RATE = 0.12
               WHEN INCOME <= 100525
                   COMPUTE TAX-RATE = 0.22
               WHEN INCOME <= 191950
                   COMPUTE TAX-RATE = 0.24
               WHEN OTHER
                   COMPUTE TAX-RATE = 0.32
           END-EVALUATE
           COMPUTE TAX-AMT = INCOME * TAX-RATE
           STOP RUN.
\end{lstlisting}

Listing~\ref{lst:python-buggy} is a translation of this program in which the boundary operator \texttt{<=} has been replaced with strict inequality \texttt{<}. It is written by hand to illustrate the class of error; it belongs to the conditional-logic category of the LLM errors reported in \S\ref{sec:rq6}:

\begin{lstlisting}[style=pythonstyle,caption={Python translation with a $<$ vs.\ $\leq$ boundary error (hand-written).},label={lst:python-buggy}]
def fed_tax_calc(income):
    if income < 47150:
        tax_rate = 0.12
    elif income < 100525:
        tax_rate = 0.22
    elif income < 191950:
        tax_rate = 0.24
    else:
        tax_rate = 0.32
    tax_amt = income * tax_rate
    return tax_amt
\end{lstlisting}

This translation is \emph{almost} correct. For the vast majority of income values, both programs produce identical results. Standard unit tests with representative inputs (e.g., \$30{,}000, \$75{,}000, \$150{,}000) would pass. The divergence manifests only at exact bracket boundaries: when \texttt{INCOME} equals exactly 47{,}150, 100{,}525, or 191{,}950.

Both programs are encoded as Z3 formulas over a shared symbolic variable \texttt{INCOME}, and the solver is asked for an assignment on which a COBOL output differs from the corresponding Python output. Z3 returns \texttt{SAT} with a concrete counterexample:

\begin{lstlisting}[style=z3style,caption={Solver result for Listings~\ref{lst:cobol-tax} and~\ref{lst:python-buggy}, condensed from the tool's output. Inputs and command: \texttt{papers/motivating/}.},label={lst:z3-output}]
TAX-AMT   SAT  counterexample INCOME = 191950
          COBOL = 46068   Python = 61424
          difference = -15356
TAX-RATE  SAT  counterexample INCOME = 191950
          COBOL = 0.24    Python = 0.32
\end{lstlisting}

At \texttt{INCOME = 191{,}950}, the COBOL program applies the 24\% rate (because $191{,}950 \leq 191{,}950$ is true), while the Python translation applies 32\% (because $191{,}950 < 191{,}950$ is false, falling through to the \texttt{else} branch). On that input the two programs differ by 15{,}356 in \texttt{TAX-AMT}.

The example shows three properties of the class of errors \textsc{Ironproof} targets: the change is syntactically minimal (one character); it changes the computed result; and tests that do not target the boundary values do not reveal it.

When the translation comes from an LLM, the correction loop (\S\ref{sec:feedback}) can send this counterexample back to the model.


\section{Approach}\label{sec:approach}

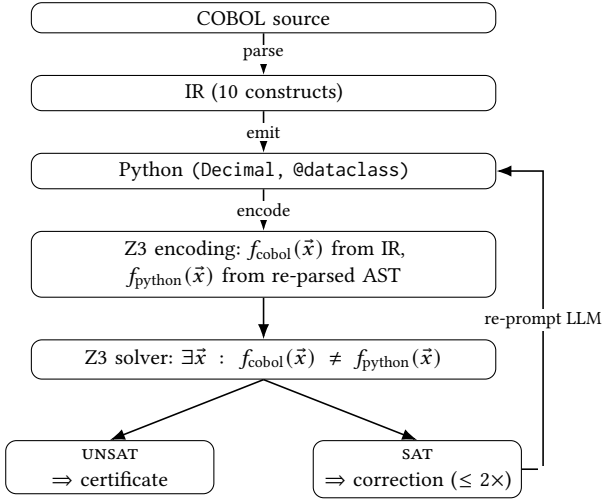
\begin{figure}[t]
\centering
\begin{tikzpicture}[
  node distance=5.5mm,
  box/.style={draw, rounded corners, align=center, font=\small,
              inner sep=3pt, text width=0.70\columnwidth},
  lbl/.style={font=\footnotesize, fill=white, inner sep=1pt},
  arr/.style={-{Latex[length=2mm]}, semithick}]
  \node[box] (cobol) {COBOL source};
  \node[box, below=of cobol] (ir) {IR (10 constructs)};
  \node[box, below=of ir] (py) {Python \texttt{(Decimal, @dataclass)}};
  \node[box, below=of py] (enc) {Z3 encoding: $f_{\text{cobol}}(\vec{x})$ from IR, $f_{\text{python}}(\vec{x})$ from re-parsed AST};
  \node[box, below=of enc] (solve) {Z3 solver:\ $\exists\vec{x}:\, f_{\text{cobol}}(\vec{x}) \neq f_{\text{python}}(\vec{x})$};
  \node[box, below=8mm of solve, xshift=-0.24\columnwidth, text width=0.30\columnwidth] (cert) {\textsc{unsat}\\$\Rightarrow$ certificate};
  \node[box, below=8mm of solve, xshift=0.24\columnwidth, text width=0.30\columnwidth] (fb) {\textsc{sat}\\$\Rightarrow$ correction ($\le 2\times$)};
  \draw[arr] (cobol) -- node[lbl]{parse} (ir);
  \draw[arr] (ir) -- node[lbl]{emit} (py);
  \draw[arr] (py) -- node[lbl]{encode} (enc);
  \draw[arr] (enc) -- (solve);
  \draw[arr] (solve.south) -- (cert.north);
  \draw[arr] (solve.south) -- (fb.north);
  \draw[arr] (fb.east) -- ++(0.28,0) node[lbl, right, xshift=-2pt] {}
        |- node[lbl, pos=0.25] {re-prompt LLM} (py.east);
\end{tikzpicture}
\caption{Architecture of the \textsc{Ironproof} pipeline. The primary data flow proceeds top-to-bottom; the feedback path (right) activates when Z3 finds a counterexample and re-prompts the LLM with the counterexample and the incorrect Python.}
\label{fig:architecture}
\end{figure}

\textsc{Ironproof} operates as a five-stage pipeline: COBOL parsing, rule-based Python generation, Z3 semantic encoding of both programs, equivalence checking, and (optionally) LLM-driven feedback correction. Figure~\ref{fig:architecture} illustrates the architecture.

\subsection{COBOL Parser}\label{sec:parser}

\textsc{Ironproof}'s parser operates in three phases: preprocessing, lexical normalization, and structural parsing.

\paragraph{Preprocessing.} A preprocessor resolves \texttt{COPY} book directives by searching a configurable set of directories for the referenced copybook files (\texttt{.cpy}, \texttt{.cbl}, \texttt{.cob}), inlining their contents at the point of reference. This is necessary because COBOL programs in production frequently factor shared data definitions into copybooks that are included at compile time.

\paragraph{Normalization.} COBOL's fixed-format layout (columns 1--6 for sequence numbers, column 7 for continuation/comment indicators, columns 8--72 for code) is normalized: comment lines (column 7 = \texttt{*} or \texttt{/}) are removed, and the code area is extracted and uppercased. This produces a flat token stream amenable to recursive-descent parsing.

\paragraph{Structural Parsing.} The parser identifies two top-level divisions: the \textsc{Data Division}, from which it extracts variable declarations with PIC clauses, and the \textsc{Procedure Division}, which it decomposes into named paragraphs and a main body. The \texttt{PicType} system models COBOL's picture-string type notation, extracting integer digit count, decimal digit count, signedness, and alphanumeric status from PIC strings such as \texttt{S9(8)V99} or \texttt{X(20)}.

The parser recognizes 10 COBOL construct types, each mapped to a typed intermediate representation (IR) node:

\begin{enumerate}
  \item \textbf{COMPUTE}: Arithmetic assignment (\texttt{COMPUTE X = expr}).
  \item \textbf{IF/ELSE}: Conditional branching with compound conditions (AND/OR), nested bodies, and optional ELSE.
  \item \textbf{EVALUATE}: Multi-way branching (COBOL's switch/case), including \texttt{EVALUATE TRUE} with arbitrary conditions per \texttt{WHEN} clause.
  \item \textbf{PERFORM}: Loop constructs, including \texttt{PERFORM~n~TIMES}, \texttt{PERFORM~UNTIL}, \texttt{PERFORM~VARYING}, and paragraph invocation.
  \item \textbf{MOVE}: Data movement (\texttt{MOVE src TO tgt1 tgt2}).
  \item \textbf{ADD/SUBTRACT/MULTIPLY/DIVIDE}: Arithmetic verbs with optional \texttt{GIVING} and \texttt{REMAINDER} clauses.
  \item \textbf{CALL}: Subprogram invocation (\texttt{CALL 'prog' USING var1 var2}).
  \item \textbf{EXEC SQL}: Embedded SQL with \texttt{INTO} host variable binding.
  \item \textbf{File I/O}: \texttt{OPEN}, \texttt{READ} (with \texttt{AT END}), \texttt{WRITE} (with \texttt{FROM}), and \texttt{CLOSE}.
  \item \textbf{GO TO}: Unconditional transfer of control, flagged for manual review.
\end{enumerate}

The parser produces a \texttt{IronproofProgram} structure containing the program name, a dictionary of typed variables, a dictionary of named paragraphs (each containing a list of IR statements), the main procedure body, resolved copybook references, and Boolean flags indicating the presence of SQL, file I/O, and GO TO constructs.

\subsection{Python Emission}\label{sec:codegen}

The code generator maps the COBOL IR to Python. Three design decisions merit discussion.

\paragraph{Decimal Arithmetic.} COBOL's fixed-point decimal arithmetic is semantically distinct from IEEE 754 floating-point. A COBOL variable declared as \texttt{PIC 9(8)V99} represents a decimal number with 8 integer digits and 2 fractional digits, with exact decimal semantics. We map all numeric COBOL variables to Python's \texttt{decimal.Decimal} type, preserving exact arithmetic. This avoids an entire class of translation bugs related to floating-point rounding.

\paragraph{Working Storage as Dataclass.} COBOL's \texttt{WORKING-STORAGE SECTION} defines program-scoped mutable state. We represent this as a Python \texttt{@dataclass} named \texttt{WorkingStorage}, with each COBOL variable mapped to a typed field with an appropriate default value derived from its PIC clause. This provides a single, typed state container that is passed to all generated functions, mirroring COBOL's flat variable namespace.

\paragraph{Paragraphs as Functions.} Each named COBOL paragraph becomes a Python function that accepts a \texttt{WorkingStorage} instance. \texttt{PERFORM paragraph-name} translates to a function call. \texttt{GO TO} statements are translated to function calls with a review comment, as COBOL's unstructured control flow cannot always be faithfully represented in Python's structured control model.

\subsection{Z3 Semantic Encoding}\label{sec:z3encoding}

The core of \textsc{Ironproof}'s verification is a dual encoding: both the COBOL program (via its IR) and the Python translation (via its AST) are mapped to Z3 formulas over shared symbolic variables.

\paragraph{Variable Mapping.} For each numeric variable in the COBOL program's \texttt{WORKING-STORAGE}, we create a Z3 \texttt{Real} variable. Variables that are read before they are written and are not fixed by a \texttt{VALUE} clause become \emph{input variables}; variables that are assigned become \emph{output variables}. The equivalence question is then: for all valuations of input variables, do the COBOL and Python programs produce identical output variables?

\paragraph{COBOL Encoding.} The COBOL IR is traversed, and each statement is interpreted symbolically:
\begin{itemize}
  \item \textbf{COMPUTE} and arithmetic verbs update the symbolic environment by evaluating the expression tree over Z3 terms.
  \item \textbf{IF} and \textbf{EVALUATE} produce Z3 \texttt{If(cond, then\_val, else\_val)} terms, merging the symbolic state from each branch.
  \item \textbf{MOVE} propagates the source value (or a Z3 constant for literals) to the target variables.
  \item Conditions are mapped to Z3 comparisons: \texttt{<=}, \texttt{<}, \texttt{>=}, \texttt{>}, \texttt{=}, and \texttt{NOT=}.
\end{itemize}

\paragraph{Python Encoding.} The Python source is parsed using Python's \texttt{ast} module. The AST is then traversed with a symbolic interpreter that maps assignments to Z3 updates, conditionals to Z3 \texttt{If} terms, arithmetic to Z3 arithmetic, and \texttt{Decimal} constructors to Z3 \texttt{RealVal} constants. The traversal produces a set of output variable formulas analogous to those from the COBOL encoding.

\paragraph{Equivalence Query.} Given COBOL output formulas $\{o^C_1, \ldots, o^C_k\}$ and Python output formulas $\{o^P_1, \ldots, o^P_k\}$ over shared input variables $\vec{x} = (x_1, \ldots, x_n)$, we assert:
\begin{equation}\label{eq:equiv}
\exists \vec{x} \in \mathcal{D} : \bigvee_{i=1}^{k} o^C_i(\vec{x}) \neq o^P_i(\vec{x})
\end{equation}
where $\mathcal{D}$ defines input bounds (e.g., $0 \leq x_j \leq 10^7$ for financial amounts). If Z3 returns \texttt{UNSAT}, no such input exists and the programs are semantically equivalent over $\mathcal{D}$. If Z3 returns \texttt{SAT}, the model provides a concrete counterexample---specific input values that produce different outputs.

\paragraph{Scope and Limitations.} This encoding covers \emph{pure computational sections}: straight-line arithmetic, conditionals, and multi-way branches without loops, I/O, or external calls. The use of Z3 \texttt{Real} (infinite-precision rationals) means the encoding is an over-approximation: it does not model overflow, truncation on store, or rounding (see \S\ref{sec:discussion}). For PERFORM loops we do not encode the loop body symbolically (which would require loop invariant synthesis); an output that depends on a loop is left unverified, unless the parser has flattened the loop identically on both sides (\S\ref{sec:noncircular}). Similarly, EXEC SQL, file I/O, CALL, and GO TO constructs are flagged but not encoded. The certificate records which constructs were detected and which were formally verified.

\subsection{LLM-Z3 Feedback Loop}\label{sec:feedback}

When Z3 returns \texttt{SAT}, \textsc{Ironproof} extracts the counterexample and uses it to construct a targeted correction prompt for the LLM. Algorithm~\ref{alg:feedback} describes this process.

\begin{algorithm}[t]
\caption{LLM-Z3 Feedback Loop}\label{alg:feedback}
\begin{algorithmic}[1]
\Require COBOL source $S_C$, maximum number of LLM calls $K=3$
\Ensure Equivalence certificate, or a report that no certificate was emitted
\State $S_P \gets \textsc{LLM-Translate}(S_C)$ \Comment{LLM call 1}
\For{$i = 1$ \textbf{to} $K$}
  \State $r \gets \textsc{Z3-Check}(\exists \vec{x}: \phi_C(\vec{x}) \neq \phi_P(\vec{x}))$
  \If{$r = \texttt{UNSAT}$ for every output}
    \State \Return $\textsc{Emit-Certificate}(S_C, S_P)$
  \ElsIf{$i = K$}
    \State \Return $\textsc{Report}(r)$
  \EndIf
  \State $S_P \gets \textsc{LLM-Correct}(S_C, S_P, \textsc{Counterexamples}(r))$ \Comment{LLM call $i+1$; the list may be empty}
\EndFor
\end{algorithmic}
\end{algorithm}

The correction prompt includes: (1) the original COBOL source, (2) the current (incorrect) Python translation, (3) the concrete counterexample input values, (4) the expected output (from the COBOL Z3 formula) and the actual output (from the Python Z3 formula), and (5) an instruction to fix the specific semantic divergence. This provides the LLM with a precise, actionable error signal rather than a vague instruction to ``try again.''

The loop makes at most $K=3$ LLM calls: one translation and up to two corrections. \textbf{We do not evaluate the loop in this paper.} An earlier evaluation on five programs was not recorded in a form we can re-derive. The current verdict rules also require that a translation be executed (a smoke run) before any certificate is emitted. A translation in the function form an LLM produces is not executed by that smoke run, so the current loop withholds a certificate on it.


\section{Implementation}\label{sec:implementation}

\textsc{Ironproof} is implemented in about 3{,}300 non-blank, non-comment lines of Python 3 (\texttt{ironproof\_core.py}). The implementation depends on three external components:

\begin{itemize}
  \item \textbf{Z3 Solver} (\texttt{z3-solver} Python package, version 4.16; other versions can change which outputs are Z3 expressions, and so the counts): Microsoft Research's SMT solver~\cite{demoura2008z3}, used for all equivalence queries. Z3 is invoked via its Python API with a configurable timeout (default: 15 seconds per variable per program).

  \item \textbf{LLM Backend}: The LLM backend for COBOL-to-Python translation and correction. We use \texttt{claude-\allowbreak opus-\allowbreak 4-7} with a 2{,}048-token output limit; all LLM results reported here were produced with that model, and the raw result files record it. The translation prompt instructs the model to produce a single function, use \texttt{Decimal} types, preserve exact conditions, and emit no comments or imports. We report results with a single model; comparative evaluation across LLMs is left to future work.

  \item \textbf{Python \texttt{ast} module}: Used to parse the LLM-generated Python code into an abstract syntax tree for Z3 encoding. This is part of the Python standard library.
\end{itemize}

\paragraph{Certificate Format.} Upon successful equivalence verification, \textsc{Ironproof} emits a JSON certificate containing: the program name, a verdict (\texttt{EQUIVALENT}, \texttt{NON-\allowbreak EQUIVALENT}, \texttt{TRADUCTION-\allowbreak INVALIDE} for a translation that does not run, or \texttt{ABSTENTION} when nothing is refuted but not every output is proved), SHA-256 hashes of both source and target, the proof method, a formal proof statement ($\neg\exists \vec{x}: f_C(\vec{x}) \neq f_P(\vec{x})$ for equivalence), a list of detected constructs (SQL, file I/O, GO TO, paragraph count, resolved copybooks), and per-variable proof results including status, timing, counterexample (if any), and details. To make ``machine-checkable'' literal rather than nominal, each proved variable also carries the self-contained SMT-LIB2 obligation Z3 discharged (declarations, assertions, and \texttt{check-sat}); a third party re-checks a certificate by feeding each obligation to any SMT solver and confirming \texttt{unsat}, without trusting our solver invocation. We verify this end-to-end in \texttt{papers/verify\_smt2.py}, which re-checks every obligation in a fresh z3 instance (113 of 113 obligations \texttt{unsat} on the first 40 proved programs).

\paragraph{Parser Coverage.} The COBOL parser handles a subset of the COBOL-85 standard. It does not handle: \texttt{INSPECT}\slash\texttt{STRING}\slash\texttt{UNSTRING} (string manipulation), \texttt{SORT}\slash\texttt{MERGE} (file sorting), \texttt{ALTER} (self-modifying code), \texttt{REDEFINES} (storage overlay), \texttt{OCCURS}\slash\texttt{OCCURS DEPENDING ON} (table handling), \texttt{COMP-3} (packed decimal), level-88 condition names (Boolean flags), or nested programs. These are recognized and flagged rather than silently ignored. We note that several of these unsupported constructs (level-88, \texttt{OCCURS}, \texttt{REDEFINES}) are pervasive in production COBOL and represent a significant gap in coverage.

\paragraph{Reproducibility.} Our source code, the scripts that produce the figures of \S\ref{sec:evaluation}, and the benchmark programs we may redistribute are at \url{https://github.com/dom-omg/ironproof-cobol}: the generated and hand-written programs, the NIST CCVS85 suite (public domain), and the community programs (MIT and CC-BY-4.0). Not reproducible from it: the runtime confrontation of \S\ref{sec:oracle}, whose harness was not versioned; the CardDemo/GenApp measurement of \S\ref{sec:discussion}, which depends on a separate pipeline; the extraction of the GnuCOBOL programs from the upstream test suite, whose files can be checked against the hashes but not regenerated by a script; and Table~\ref{tab:refuted}, a dated record of an earlier pipeline. The other two sources are not redistributed. The GnuCOBOL test suite is GPL-3.0 and available upstream; the dscobol collection carries no license. Both are identified file by file by SHA-256 in \texttt{benchmark/CORPUS\_TIERS.sha256}, so a replication can check that it runs on the exact bytes we measured. The tool can be invoked via command line:

\begin{lstlisting}[basicstyle=\ttfamily\scriptsize,frame=single,float=tbp]
$ PYTHONHASHSEED=0 python3 ironproof_core.py --cobol prog.cbl
$ PYTHONHASHSEED=0 python3 ironproof_core.py --demo-bug
\end{lstlisting}


\section{Evaluation}\label{sec:evaluation}

We evaluate \textsc{Ironproof} along five research questions:

\begin{description}
  \item[RQ1] On what fraction of COBOL programs do parsing and Python emission run to completion without crashing?
  \item[RQ2] How many translations does Z3 prove semantically equivalent?
  \item[RQ3] What is the runtime performance of the pipeline?
  \item[RQ4] Can PIC-bounded overflow detection identify errors missed by unbounded verification?
  \item[RQ5] How does \textsc{Ironproof}'s rule-based transpilation compare to LLM-only translation?
\end{description}

\subsection{Benchmark}\label{sec:benchmark}

We assembled a benchmark suite of 2{,}345 COBOL files from multiple open-source repositories (Table~\ref{tab:benchmark}). They include 76 byte-identical copies, in 59 groups, mostly among the GnuCOBOL tests; we count files, as every figure of this paper does. The suite comprises: (1) 1{,}632 programs extracted from the GnuCOBOL compiler test suite, which exercises the full breadth of COBOL-85 constructs; (2) 69 programs from the NIST CCVS85 standard validation suite~\cite{nist1985cobol} (public domain); (3) 113 programs from the dscobol enterprise COBOL collection (DB2, VSAM, batch); (4) 31 community programs: 23 from COBOL-Examples (MIT) and 8 from the Open Mainframe Project COBOL Programming Course (OMP, CC-BY-4.0); (5) 60 hand-written programs covering all 10 supported construct types plus 5 stress tests targeting DIVIDE INTO semantics, PIC overflow, and decimal truncation; and (6) 440 systematically generated programs covering combinatorial variations of arithmetic, conditionals, and control flow.

\begin{table}[t]
\centering
\caption{Benchmark composition: 2{,}345 COBOL programs.\protect\footnotemark}
\label{tab:benchmark}
\small
\setlength{\tabcolsep}{2.5pt}
\begin{tabular}{@{}llr@{}}
\toprule
\textbf{Source} & \textbf{Description} & \textbf{Retained} \\
\midrule
GnuCOBOL tests      & Compiler test suite (GPL-3.0)                 & 1{,}632 \\
Generated           & Combinatorial construct variations             & 440 \\
dscobol collection  & Enterprise COBOL (DB2, VSAM, batch)           & 113 \\
NIST CCVS85         & Standard validation suite (public domain)     & 69 \\
Hand-written        & All 10 construct types + 5 stress tests        & 60 \\
Community / OMP     & Open-source programs (MIT, CC-BY-4.0)          & 31 \\
\midrule
\textbf{Total}      &                                               & 2{,}345 \\
\midrule
\multicolumn{2}{l}{\emph{Enter checking path (COBOL yields an encodable output)}} & 782 \\
\multicolumn{2}{l}{\emph{\quad of which proved equivalent}} & 606 \\
\multicolumn{2}{l}{\emph{No encodable output (I/O, SQL, loops, INSPECT)}} & 1{,}563 \\
\bottomrule
\end{tabular}
\end{table}
\footnotetext{The initial GnuCOBOL extraction yielded 2{,}072 candidate programs; we excluded 440 that exercise advanced language features (intrinsic functions, reference modification, SCREEN SECTION, inter-program communication) producing Python code patterns our Z3 AST encoder cannot handle. The NIST CCVS85 suite contains approximately 459 test programs; we retained 69 that parsed successfully under our recursive-descent parser---the remaining programs exercise COBOL features outside our parser scope (INSPECT, REDEFINES, OCCURS, level-88 conditions, nested programs). These exclusions reflect limitations of our implementation, not of the source COBOL. All dscobol, community, hand-written, and generated programs were retained without exclusion.}

Counting non-empty lines, programs range from 4 to 7{,}693 lines (median: 18, mean: 68). The proved fragment (606 programs) ranges from 8 to 7{,}693 lines (median: 17, mean: 130); a few very large GnuCOBOL test programs pull the mean up. The GnuCOBOL, NIST, dscobol, and community programs are \emph{independently authored} COBOL: not written by the author of this paper. They are not free of selection, since the exclusions in the footnote to Table~\ref{tab:benchmark} remove programs outside our parser's or encoder's reach; only the generated and hand-written sets are ours. We use this definition of ``independently authored'' throughout. The majority (1{,}563 programs, 66.7\%) yield no Z3-encodable output at all; a further 176 enter the checking path without being proved (101 partially verified, 75 pipeline failures). We keep these categories distinct in \S\ref{sec:rq2} rather than reporting a single ``outside scope'' figure, since the last group consists of our own pipeline's failures rather than properties of the input.

\subsection{Results: RQ1 -- Parse and Emission Crash-Freedom}\label{sec:rq1}

Table~\ref{tab:rq1} reports the parse and emission rates by source.

\begin{table}[t]
\centering
\caption{Parse and Python emission by source. Emission below 100\% is a refusal by our generator, not a crash.}
\label{tab:rq1}
\small
\begin{tabular}{lrrr}
\toprule
\textbf{Source} & \textbf{N} & \textbf{Parse} & \textbf{Python emission} \\
\midrule
GnuCOBOL tests       & 1{,}632 & 100\% & 79.2\% \\
Generated            & 440    & 100\% & 100\% \\
dscobol collection   & 113    & 100\% & 70.8\% \\
NIST CCVS85          & 69     & 100\% & 79.7\% \\
Hand-written + stress & 60    & 100\% & 95.0\% \\
Community / OMP      & 31     & 100\% & 67.7\% \\
\midrule
\textbf{Total}       & 2{,}345 & 100\% & 82.9\% \\
\bottomrule
\end{tabular}
\end{table}

The parser completes on all 2{,}345 benchmark programs; for the NIST suite this holds by selection, since the footnote to Table~\ref{tab:benchmark} retains only programs that parsed. Python emission completes on 1{,}945 (82.9\%); on the other 400 our generator refuses to emit. In 276 of them a COBOL name cannot be mapped to a Python identifier, and in 124 the emitted text would not compile as Python. An earlier version of the generator emitted text in both cases and counted it as a success. Where a program rests on unsupported constructs---INSPECT, STRING, OCCURS, REDEFINES, nested programs---those sections are recognized but not modeled, and level-88 condition names are skipped by the parser. Whether a program enters the checking path is decided on the COBOL side, before emission (\S\ref{sec:rq2}). A refusal therefore counts against the proof rate only when the program's COBOL yields something encodable: 47 of the 400 do, and they are among the 75 pipeline failures.

\paragraph{Emission is a refusal point, not a proof.} That emission completes still does not mean the emitted Python is right. It means the text compiles and every name it writes is a declared data name. Fitness for verification is decided downstream (\S\ref{sec:rq2}).

\subsection{Results: RQ2 -- Z3 Proof Rate}\label{sec:rq2}

A program \emph{enters the checking path} when its COBOL side yields at least one Z3-encodable output variable. This is decided before Python emission and does not depend on our generator. Of the 2{,}345 programs, 782 enter; the remaining 1{,}563 yield none and are flagged as outside the encodable fragment. A parse or a COBOL encoding that raises keeps the program in the denominator, since we cannot tell that it is out of scope and the failure is ours. Of the 782, 707 complete a decision procedure (606 proved equivalent, 101 partially verified, none refuted) and 75 fail inside our pipeline. Table~\ref{tab:rq2} summarizes the outcomes.

\begin{table}[t]
\centering
\caption{Z3 equivalence outcomes on 2{,}345 programs. Every program falls in exactly one row; no outcome is folded into another.}
\label{tab:rq2}
\small
\begin{tabular}{lr}
\toprule
\textbf{Outcome} & \textbf{Count (\%)} \\
\midrule
UNSAT (equivalent, proof emitted)          & 606 (25.8\%) \\
SAT (counterexample found)                 & 0 (0.0\%) \\
Partial (some outputs not verified)        & 101 (4.3\%) \\
Pipeline failure (not checked)             & 75 (3.2\%) \\
No Z3-encodable output (I/O, SQL, loops)   & 1{,}563 (66.7\%) \\
\midrule
\textbf{Total}                             & 2{,}345 (100\%) \\
\bottomrule
\end{tabular}
\end{table}

\paragraph{Proof rate.} Among the 782 programs entering the checking path, 606 (\textbf{77.5\%}) are proved equivalent. Restricted to the 707 programs on which a decision procedure terminates, the rate is 85.7\%. We report the former as the headline figure because the latter excludes the 75 programs our own pipeline could not process, and excluding one's own failures from a denominator makes the resulting rate unable to fall---a rate that cannot go down does not measure anything. A solver query that times out (15 seconds per output variable) returns \texttt{UNKNOWN}, and its program is counted as partial, never as proved.

\paragraph{What the 101 partial verdicts are.} In 89 of them, every output that is not proved is marked unverified: the encoder could not carry its value through an unmodeled construct upstream, and an output we did not encode is not reported as equivalent. In 8, the Python side produces no value for an output the COBOL side defines. In 2, the model is flagged as corrupted. In the last 2, Z3 finds a counterexample whose witness assigns inputs that have no declared range; we do not report a counterexample we cannot place in the program's input domain as a refutation.

\paragraph{What the proofs quantify over.} The obligation $\neg\exists \vec{x}: f_C(\vec{x}) \neq f_P(\vec{x})$ ranges over the program's free inputs: variables read before they are written and not fixed by a \texttt{VALUE} clause. In this benchmark that domain is usually empty, and what that means depends on whether data can enter the program at all. Of the 606 proved programs, \textbf{540 are closed}: they have no input surface (no file, \texttt{ACCEPT}, \texttt{LINKAGE} parameter, embedded SQL or \texttt{SORT}) and a \texttt{VALUE} clause fixes every variable. A closed program has a single execution, so evaluating both artifacts on it covers the program's whole behaviour. The proof is complete, but it is a proof about one computation, and 453 of these 540 are programs we generated or wrote. \textbf{37 are open} yet carry no free input: they read a file, accept a value or receive parameters, and the encoder models none of those inputs as free. For them the proof covers one execution of a program that has many; all 37 are independently authored. The remaining 29 have free inputs, and \textbf{14} have a proved output that depends on one. Only for those 14 does the quantifier range over a domain that matters. Each certificate records its domain (\texttt{input\_domain}). The input-surface detector is a keyword match on code columns, not a parser, and can miss an unusual form; the counts are re-derivable with \texttt{papers/rq\_breakdown.py}.

\paragraph{Reproducibility: deterministic by construction.} An earlier version of the pipeline was not deterministic: three runs on an unchanged tree and corpus produced 586, 587, and 588 proved programs and either six or seven refutations. The cause was three set-union iterations (\texttt{set(then\_env) | set(else\_env)}) in the Z3 encoder---Python randomizes \emph{set} iteration order per process, which changed the order in which the merged \texttt{If}-expressions were assembled and thus the order assertions reached the solver, flipping a small number of marginal queries between \texttt{unsat}, \texttt{sat}, and \texttt{unknown}. (An earlier note attributing this to \emph{dict} iteration was wrong: Python~3.7+ dicts are insertion-ordered and unaffected.) Sorting those unions by variable name removed the dependency at its source. The figures of this version were produced with \texttt{PYTHONHASHSEED=0}; the two scripts that check them against this paper (\texttt{check\_paper\_numbers.py}, \texttt{check\_paper\_consistency.py}) refuse to run without it. A second defect of the same encoder made the figures unreproducible in a different way: the merge rebuilt \texttt{If(c, x, x)} for every assigned variable, even when no branch had touched it, and on one 7{,}693-line GnuCOBOL program the term grew until a run no longer finished. The merge now keeps \texttt{x}, which has the same value; one consequence is that six outputs previously dropped as unverified, because \texttt{x} sat under a contaminated condition, are now encoded.

\paragraph{Composition of the proved fragment.} The 606 proved programs span all six benchmark sources: 440 generated construct variations, 126 GnuCOBOL compiler tests, 15 dscobol enterprise programs, 13 hand-written programs, 10 NIST CCVS85 programs, and 2 community programs. Generated and hand-written programs make up 74.8\% of them, which reflects the encodable scope: most independently authored COBOL uses file I/O, loops, or string operations outside the current encoding.

\paragraph{Provenance of the proof rate.} A self-authored benchmark can be written to favour the tool, so we separate the proof rate by who wrote the code, over the \emph{full checking-path denominator}, including our own failures. Of the 291 \emph{independently authored} programs that enter the checking path (GnuCOBOL, NIST CCVS85, dscobol, and community sources---not written by us), \textbf{153 prove equivalent (52.6\%)}; none is refuted, 66 are partially verified, and 72 fail inside our pipeline. \emph{72 of the 75 pipeline failures fall on this independent code}, where real-world constructs exceed what the current encoding and generator handle. On the 491 programs we generated or wrote, 453 prove (92.3\%), 35 are partially verified and 3 fail. That rate reflects programs that stay inside the encoder's designed scope, and all 453 of those proofs are over closed programs; this is why we do not headline it. On the independent side, 37 of the 153 proofs cover a single execution of an open program (\S\ref{sec:rq2}); counting only proofs that are complete for their program, the independent rate is 116 of 291 (39.9\%). The figure for code we did not write is therefore 52.6\%, reported without excluding our failures from its denominator; the 77.5\% of \S\ref{sec:rq2} is a blend of the two populations and should be read as such. The split is re-derivable with \texttt{papers/provenance\_breakdown.py}. \S\ref{sec:oracle} reports what a reference COBOL implementation said about an earlier version of this figure.

\paragraph{The 75 pipeline failures are ours, not the source COBOL's.} For 27 programs, encoding the COBOL side raises a Z3 exception. For 47 programs whose COBOL yields an encodable output, our generator refuses to emit: 28 contain a name it cannot map to a Python identifier and 19 would yield Python that does not compile. One more fails while encoding or solving the Python side. None of these is a program whose COBOL constructs lie outside the fragment, and we report them separately so that the pipeline's own defects are visible rather than absorbed into a statement about the input language. The classification of every program is defined once, in \texttt{papers/classify.py}, and imported by every script that produces a figure of this paper.

\subsection{What the Proof Rate Does and Does Not Establish}\label{sec:noncircular}

A natural objection to any self-verifying transpiler is circularity: if both sides of the equivalence query are derived from the same intermediate representation, the solver is being asked whether a system agrees with itself, and \texttt{UNSAT} is guaranteed regardless of correctness. We state precisely where the two sides come from, and then give the empirical evidence that they are not the same side.

The COBOL side of each query is built from the IR produced by the parser. The Python side is \emph{not}: it is obtained by re-parsing the emitted Python source text with the standard \texttt{ast} module and encoding the resulting syntax tree independently (\S\ref{sec:z3encoding}). The two encoders therefore share the parser but not the code generator, and a query can distinguish them exactly when the generator's output diverges from the IR it was generated from.

\paragraph{Evidence that the query can fail.} If the encoding were tautological, \texttt{SAT} would be unreachable. On the current code no benchmark program is refuted, so the benchmark no longer supplies that evidence by itself; two other recorded sources do. In the April 2026 baseline (\S\ref{sec:rq6}), Z3 returned \texttt{SAT} with a concrete counterexample on 38 LLM translations. And on 2026-08-26 our own generator was refuted on six programs (Table~\ref{tab:refuted}): in \texttt{gnucobol\_320}, Z3 returned a counterexample on \texttt{CNTR-RCD-WRITTEN} in which the COBOL side evaluated to $0$ and the emitted Python to $0+1$. Those six are not refuted today, but not because the defects were repaired. The later hardening stops them earlier: three are refused by the generator, two no longer yield an encodable output, and one is partially verified. We report this rather than present the zero as a cleaner generator.

\begin{table}[t]
\centering
\caption{The programs on which Z3 refuted equivalence between the COBOL IR and the generated Python on 2026-08-26, each a code-generation defect found by the verifier, and how the current pipeline classifies them.}
\label{tab:refuted}
\small
\begin{tabular}{@{}llr@{}}
\toprule
\textbf{Program} & \textbf{Refuted output(s)} & \textbf{Now} \\
\midrule
\texttt{dscobol\_016}  & \texttt{FINALSTOCKTOTAL}, \texttt{PRNVALUE} & partial \\
\texttt{dscobol\_031} & \texttt{RECORDS-READ} & refused \\
\texttt{gnucobol\_320}          & \texttt{CNTR-RCD-WRITTEN} & no output \\
\texttt{gnucobol\_321}          & \texttt{CNTR-RCD-WRITTEN} & no output \\
\texttt{gnucobol\_322}          & \texttt{CNTR-RCD-WRITTEN}, \texttt{HEAP-ID} & refused \\
\texttt{gnucobol\_1490}         & \texttt{SUB1} & refused \\
\bottomrule
\end{tabular}
\end{table}

\paragraph{What remains outside the guarantee.} Because both sides inherit the same parser, the obligation establishes fidelity of \emph{emission} (IR $\to$ Python), not fidelity of \emph{parsing} (COBOL $\to$ IR). A construct the parser misreads is misread identically on both sides, and the query returns \texttt{UNSAT} on a translation that is wrong with respect to the COBOL standard. A proof emitted by \textsc{Ironproof} should therefore be read as: \emph{the generated Python computes what our IR says the COBOL computes}. Establishing that the IR itself says the right thing requires an independent oracle; \S\ref{sec:oracle} reports what such an oracle says when we build one.

A second gap is narrower and we state it explicitly because it is easy to miss: the Python side of the obligation is a \emph{model} of the emitted file, not the file as CPython runs it. \texttt{\_build\_python\_z3} enumerates every function definition in the emitted module and encodes each body in source order against a shared environment; it never consults the module's \texttt{\_\_main\_\_} block. When the emitted \texttt{\_\_main\_\_} calls a different entry point than that enumeration effectively assumes, the theorem and the artifact describe different computations. We quantify this in \S\ref{sec:oracle}. Relatedly, the Python environment is seeded with the COBOL \texttt{VALUE} constants (\texttt{\_build\_python\_z3(python\_src, all\_cobol\_vars)}), so both sides share the initial state by construction and no initialization discrepancy between the two artifacts is expressible as a counterexample.

\paragraph{How much of the proof rate is fully modeled.} This emission/parsing distinction is not uniform across the 606 proofs, so we quantify the split rather than leave it to the reader. Of the 606, \textbf{575 (94.9\%)} are \emph{fully modeled}: the encoder captures their entire control flow, with no \texttt{PERFORM ... UNTIL} loop, no file operation, and no guard the parser had to collapse to a constant. The remaining 31 contain at least one such construct---a loop, a file operation, or a condition the shared parser flattens \emph{identically on both sides}. For those 31, the equivalence query still holds, but as an emission-fidelity statement over the flattened representation rather than a claim about the program's full semantics. The breakdown is re-derivable with \texttt{papers/modeled\_breakdown.py}.

\subsection{Confronting the Proofs with a Reference Runtime}\label{sec:oracle}

The limitation stated above---that a proof is relative to our IR, and that establishing the IR itself requires an independent oracle---is the one an evaluator will press hardest. We therefore built the oracle and report what it says about our own results, including where it contradicts them.

\paragraph{Snapshot.} This confrontation ran on 2026-08-26, on the pipeline as it stood before the changes that produce the figures of \S\ref{sec:rq2}. Its harness was not versioned with the artifact and has not been re-run on the current code. The figures below therefore describe that snapshot, not the current pipeline.

\paragraph{Method.} For each program we auto-generate a \texttt{DISPLAY} of every variable in the proof's domain, insert it before each reachable terminator and at end of unit, compile the instrumented source with GnuCOBOL 3.2.0 (\texttt{cobc -x}), execute the emitted Python, and compare the two final states as exact decimals. The comparison is reported in eight distinct states so that ``not confronted'' is never conflated with ``refuted'': disagreement, agreement, no probe reached, unreadable probe, compile failure, runtime failure, external input without a fixture, and Python failure.

\paragraph{Denominator.} Of the 113 independently authored programs holding an \texttt{EQUIVALENT} verdict in that snapshot, 60 compile as written under \texttt{cobc -x}, 71 compile once modules and alternative source formats are accepted, and \textbf{49 are confrontable end to end}. The 64 that are not are dominated by compiler bug-report fragments that do not compile standalone, subprograms requiring a driver, and programs whose input surface has no fixture. This is a selection effect that works \emph{against} the reported figure: the 49 are the simplest programs in the set---self-contained, single compilation unit, no external input.

\paragraph{Result.} On those 49, \textbf{24 disagree with the reference runtime} on at least one variable in the proof's own domain. Each disagreement survives an invariance check against any rescaling of our \texttt{PICTURE} interpretation, so none is an artifact of how we read the probe output. The dominant mechanism is store-time truncation: the encoder models numeric data as unbounded reals, whereas COBOL truncates on assignment to the receiving field's \texttt{PICTURE}, applies \texttt{ROUNDED} only where declared, and leaves a field unchanged on an unhandled size error. This is the same phenomenon our PIC-bounded analysis (\S\ref{sec:rq5}) flags statically, observed here as a divergence rather than a warning.

\paragraph{Model versus artifact.} Comparing the two Python execution models directly---the enumeration the obligation encodes against the emitted \texttt{\_\_main\_\_}---\textbf{18 of the 49 diverge from each other}, independently of COBOL. On 13 of them the obligation agrees with the reference runtime while the emitted file, executed as a program, does not: the emitted \texttt{\_\_main\_\_} calls a function whose body is empty. For those programs the theorem is not false, but it does not transfer to the artifact a user would run.

\paragraph{Consequence for the headline figures.} What the oracle established is how much of the snapshot's independent proof rate survived an independent check: on the confrontable subset, 24 of 49 did not. A verdict of \texttt{EQUIVALENT} should be read as an emission-fidelity statement relative to our IR, not as a claim validated against a COBOL implementation, unless the certificate also carries a runtime confrontation. Re-running the confrontation on the current pipeline, with a versioned harness, is the first item of our future work.

\paragraph{Scope of the oracle.} The display properties the probe relies on---explicit sign, readable packed-decimal, an inserted decimal point for implied-\texttt{V} fields---are those of GnuCOBOL 3.2.0 in this configuration, not properties of COBOL: sign representation and binary truncation are dialect options. A replication on another implementation should re-establish them before comparing numbers.

\subsection{Results: RQ3 -- Runtime Performance}\label{sec:rq4}

Table~\ref{tab:rq4} reports the execution time breakdown across pipeline stages.

\begin{table}[t]
\centering
\caption{Execution time per program across pipeline stages, measured over the 606 proved programs in a single process (Apple M4, Python 3.14, Z3 4.16). Reported in milliseconds with median and 95th percentile alongside the mean: the encoding stage is strongly right-skewed, and a mean alone misrepresents it. The total mean is the sum of stage means (means are additive); a total median and p95 are not reported because medians and quantiles are not additive across stages.}
\label{tab:rq4}
\small
\begin{tabular}{lrrr}
\toprule
\textbf{Stage} & \textbf{Mean} & \textbf{Median} & \textbf{p95} \\
\midrule
COBOL parsing                    & 1.61 & 0.20 & 1.09 \\
Python emission                  & 1.10 & 0.29 & 1.02 \\
Z3 encoding (COBOL + Python)     & 2{,}797.39 & 1.55 & 14.70 \\
Z3 equivalence check             & 477.01 & 0.61 & 2.30 \\
\midrule
\textbf{Total} (mean is additive)& \textbf{3{,}277.12} & --- & --- \\
\bottomrule
\end{tabular}
\end{table}

The typical proved program completes each stage in about a millisecond: the median is 1.55\,ms for encoding and 0.61\,ms for the equivalence check. The means tell a different story, and we report both. A few very large GnuCOBOL test programs dominate them: the slowest program takes 296 seconds to encode, and the slowest equivalence check takes 36 seconds. Encoding's mean (2.8\,s) is about 1{,}800 times its median. Parsing and emission are negligible at both ends of the distribution, and the solver is not the bottleneck; constructing the terms handed to it is. The measurement run, which classifies all 2{,}345 programs and also re-runs parse and emission on each of them, took 33 minutes on one core, almost all of it on those few programs.

\subsection{Results: RQ4 -- PIC-Bounded Overflow Detection}\label{sec:rq5}

\textsc{Ironproof}'s overflow detection mode checks whether any output variable can exceed its PIC-declared range under valid inputs. This goes beyond equivalence checking: even if COBOL and Python produce identical results, an overflow in the COBOL program indicates a latent bug that the migration should surface rather than silently reproduce.

\paragraph{At benchmark scale.} We ran the detector over every program entering the checking path. Of the 782, 754 complete the overflow analysis; for 27 the COBOL encoding raises before the query, and for 1 the overflow query itself fails. The detector uses only the COBOL side, so the 47 programs our generator refuses are analyzed. \textbf{167 of the 754 (22.1\%) contain at least one output variable that Z3 can drive outside its PIC-declared range}, for 4{,}845 flagged variables in total. The distribution is concentrated: 99 programs flag a single variable, 31 flag two, and 37 flag three or more; the maximum, 500 variables, comes from one very large GnuCOBOL test program. The figures are re-derivable with \texttt{papers/overflow\_breakdown.py}.

\textbf{This is an upper bound on real risk, not a defect count.} The analysis quantifies over \emph{unconstrained} symbolic inputs, so any accumulator narrow enough will admit some valuation that overflows it---a \texttt{PIC 9(2)} record counter is flagged even if the file it counts never holds a hundred records. Establishing which flags correspond to reachable states requires input constraints the source does not carry, which we do not infer. The useful reading is a ranking of where narrow fields meet unbounded arithmetic, not a verdict on each site. What the measurement does establish is that the condition is common rather than exotic: it is not confined to a stress test we wrote.

\paragraph{Worked example.} On our overflow stress test (program \texttt{OVERFLOW-\allowbreak TEST}, 13 variables), the detector identifies three PIC range violations, each with a concrete witness:

\begin{itemize}
  \item \texttt{WS-OVERFLOW-SUM} (\texttt{PIC 9(2)}, range [0, 99]): computed value 105 exceeds capacity. In COBOL, the high-order digit is truncated, producing 05. In Python, the value is 105.
  \item \texttt{WS-TOTAL-PCT} (\texttt{PIC 9(2)V99}, range [0, 99.99]): three percentages summing to 100.00 exceed the field capacity.
  \item \texttt{WS-NEW-BAL} (\texttt{PIC 9(3)V99}, range [0, 999.99]): deposit pushes balance to 1005.50, exceeding 3-digit capacity.
\end{itemize}

For programs with symbolic inputs, the Z3 solver searches for input valuations that produce out-of-range outputs, providing concrete counterexamples when overflow is possible.

The check targets computations where COBOL's truncation on store diverges from Python's arbitrary-precision arithmetic, which the unbounded equivalence query does not model.

\subsection{Results: RQ5 -- Baseline Comparison: Rule-Based vs. LLM-Only}\label{sec:rq6}

To quantify the value of \textsc{Ironproof}'s rule-based transpilation and Z3 verification layer, we compare against a baseline consisting of direct LLM translation without any formal verification. For each of the 586 programs in the proved fragment as identified when this experiment was run, in April 2026, we prompt the same LLM used in the feedback loop (\S\ref{sec:feedback}) to translate the COBOL source to Python, then verify the LLM-generated translation against the COBOL IR using Z3. This set is not the current proved fragment (606 programs, \S\ref{sec:rq2}), and the baseline has not been re-run on the current code, which would require new LLM calls; we report the set and the encoder actually used rather than restate them against a run the baseline did not use.

\begin{table}[t]
\centering
\caption{Baseline comparison on the 586 programs that were proved at the time the baseline was run. \textsc{Ironproof}'s column is 586/586 by construction: this set was \emph{defined} as the programs it proved, so the figure characterizes the LLM, not \textsc{Ironproof}. \textsc{Ironproof}'s own failure rate is reported in Table~\ref{tab:rq2}.}
\label{tab:rq6}
\small
\setlength{\tabcolsep}{3.5pt}
\begin{tabular}{@{}lrrrr@{}}
\toprule
\textbf{Approach} & \textbf{N} & \textbf{Proved} & \textbf{Errors} & \textbf{Unverif.} \\
\midrule
\textsc{Ironproof} (rule + Z3) & 586 & 586 (100\%)  & 0          & 0 \\
LLM-only (no verif.)        & 586 & 520 (88.7\%) & 38 (6.5\%) & 28 (4.8\%) \\
\bottomrule
\end{tabular}
\end{table}

Table~\ref{tab:rq6} summarizes the results. \textsc{Ironproof}'s column is 586 of 586 by construction (the set was defined as the programs it proved). The LLM-only baseline produces 38 confirmed semantic errors (Z3 returns SAT with a concrete counterexample) and 28 translations that fall outside our verification scope---either because the LLM emitted Python constructs our Z3 encoder cannot model, or because the equivalence query timed out as UNKNOWN. On this set \textsc{Ironproof} has neither category, because the set was defined as the programs it proved. We report the 38 SAT-confirmed errors and the 28 unverifiable translations together (66 programs total, 11.3\%) as the gap that formal verification detects; only the 38 SAT cases are categorized in the error taxonomy below, since UNKNOWN outcomes do not yield a diagnostic counterexample.

\paragraph{Where the baseline gap actually is.} This 586-program set is the fragment \textsc{Ironproof} proved, so its 100\% column is true by construction and characterizes the LLM, not \textsc{Ironproof}---whose own rate on independent code is 52.6\% today (\S\ref{sec:rq2}). The LLM's failures concentrate sharply by provenance: of the 94 independently authored programs in this baseline subset (GnuCOBOL, NIST, dscobol) it proves only 40 (42.6\%), commits 30 Z3-confirmed semantic errors, and leaves 24 unverifiable, whereas on the 490 programs we generated or wrote it proves 480 (98.0\%). The unverifiable class is a limitation of the shared encoder, not a difference between the generators---the LLM's 24 independent unverifiable cases are the same kind of gap as \textsc{Ironproof}'s own independent pipeline failures---so the encoder-neutral comparison looks only at Z3-decided outcomes. There the LLM was refuted on 30 of 94 (32\%) in April. For the rule-based generator we have no measurement from April on the same population. On 2026-08-26 it was refuted on 6 programs (Table~\ref{tab:refuted}), and on the current code on none of its 291 independent checking-path programs (153 proved, 66 partially verified, 72 pipeline failures). The encoder changed between these dates, so we do not set either figure against the LLM's 32\%. The split is re-derivable with \texttt{papers/baseline\_split.py}.

\paragraph{Error distribution by source.} The LLM failure rate varies sharply across benchmark sources, ranging from 0\% on synthetic inputs to 80\% on enterprise code. On the 440 generated programs (pure arithmetic and conditionals), the LLM achieves 100\% correctness---unsurprising, since these programs exercise narrow, predictable patterns. On independently authored code, failures dominate: \textbf{27 of 53} GnuCOBOL compiler tests fail (51\% failure rate), \textbf{7 of 16} NIST CCVS85 programs fail (44\%), and \textbf{20 of 25} dscobol enterprise programs fail (80\%). Aggregated across the GnuCOBOL, NIST, and dscobol sources---94 programs, the subset of independently authored code used for this baseline---the LLM fails on \textbf{54 programs (57\%)}---30 Z3-confirmed semantic errors and 24 unverifiable translations, not 54 confirmed errors. (The few community programs, also independently authored, are counted with the hand-written set for this comparison rather than in the 94.) The remaining 12 failures occur on the 50 hand-written and 2 community programs (23\% combined failure rate). On this baseline, LLM translation failed far more often on independently authored COBOL than on the programs we generated.

\paragraph{Error taxonomy.} Z3 counterexamples on the 38 failed LLM translations reveal semantic errors across three categories:

\begin{itemize}
  \item \textbf{Initialization and counting errors} (21 programs): The LLM produces incorrect initial values or off-by-one counter updates, causing divergences on the first or last loop iteration. These errors are characteristic of programs with file I/O record counting and paragraph-level control flow.

  \item \textbf{Conditional logic errors} (10 programs): The LLM misinterprets compound conditions (\texttt{AND}/\texttt{OR} combinations), nested \texttt{EVALUATE} branches, or \texttt{NOT} conditions, producing incorrect branch selection for specific input ranges.

  \item \textbf{Arithmetic and type errors} (7 programs): The LLM introduces computation errors including decimal precision loss, incorrect operator precedence in chained arithmetic, and missing \texttt{GIVING} clause handling.
\end{itemize}

On the independently authored COBOL in our baseline, LLM-only translation failed on 57\% of programs. The rule-based code generator is not defect-free either: in the 2026-08-26 snapshot Z3 refuted 6 of its translations (\S\ref{sec:noncircular}). What the verification layer provides is not a generator that cannot err, but a machine-checkable filter that separates the translations we can stand behind from those we cannot. For programs outside the Z3-encodable fragment, \textsc{Ironproof} makes no equivalence claim.


\section{Discussion}\label{sec:discussion}

\subsection{Threats to Validity}

\paragraph{Internal Validity.} The full benchmark spans 2{,}345 programs (median: 18 non-empty lines, mean: 68, range: 4--7{,}693). The proved fragment (606 programs) has median 17 lines, mean 130, max 7{,}693 lines. More important than size is the input domain: 37 proofs cover a single execution of a program that reads input the encoder does not model, and only 14 quantify over an input that a proved output depends on (\S\ref{sec:rq2}). The provable fragment is concentrated in pure-computation programs; larger programs use I/O, loops, and string operations outside Z3 scope. Our evaluation measures \textsc{Ironproof}'s capability on this pure-computation fragment, not on full-system migration. Production COBOL programs commonly span thousands of lines with complex inter-paragraph control flow.

The Z3 encoding uses \texttt{Real} arithmetic (infinite-precision rationals) to model both COBOL fixed-point and Python \texttt{Decimal} values. This is an \emph{over-approximation}, not a sound encoding of COBOL semantics: Z3 \texttt{Real} does not model the finite range constraints of COBOL PIC clauses (e.g., \texttt{PIC 9(2)} can only represent values 0--99; incrementing 99 causes overflow or truncation depending on the compiler), nor does it model the \texttt{ROUNDED} clause or truncation behavior. Consequently, the encoding may \emph{miss} divergences caused by overflow, truncation, or rounding---an equivalence proof does not cover divergences caused by overflow, by truncation on store (which occurs within range, for fractional digits), or by rounding. A more precise encoding using Z3 \texttt{BitVec} or bounded integer constraints with explicit range checks would be required to model these behaviors faithfully. We consider this a significant limitation that future work must address.

\paragraph{External Validity.} Our benchmark draws from the NIST CCVS85 validation suite and the GnuCOBOL compiler test suite---both independently authored and publicly available. However, these are \emph{test programs}, not production codebases. To measure the distance, we ran the same encoder on two public business-application corpora pinned at fixed commits: AWS CardDemo (44 programs, median 509 lines, Apache-2.0, commit \texttt{59cc6c2}) and IBM GenApp (31 programs, EPL-2.0, commit \texttt{f6f3f4b}). No program of either is modeled end to end. At paragraph scale, 1 of CardDemo's 573 paragraphs is modeled, a single statement, and none of GenApp's 139. The dominant obstacles are CICS identifiers (\texttt{EIBCALEN}, \texttt{EIBTRNID}), embedded SQL and \texttt{SQLCODE}, alphanumeric targets, and \texttt{GO TO}; 46 CardDemo and 5 GenApp copybooks were not found, which can only lower these counts further. Evaluation on proprietary production code from specific industries would strengthen external validity, but on public business code the current encoder's reach is close to zero.

The LLM translation quality depends on the specific model used. Results may differ with other models or future model versions. We do not claim generalizability to all LLMs, and a comparative evaluation across multiple LLMs remains necessary to establish robustness.

\paragraph{Construct Validity.} We measure ``semantic accuracy'' as Z3 equivalence on the pure computational fragment. This is a precise but narrow metric: it does not cover I/O behavior, SQL semantics, or loop termination. Programs with no encodable output are outside the checking path; programs that enter it are counted even when they contain constructs the encoder does not model (\S\ref{sec:noncircular}).

\subsection{Limitations}

\textsc{Ironproof}'s formal verification is limited to \emph{pure computational sections}---arithmetic, conditionals, and multi-way branches without loops, I/O, or external calls. This is a fundamental limitation of the encoding approach: representing loops symbolically requires loop invariant synthesis, which is itself an undecidable problem in general. We discuss directions for relaxing this limitation below.

The COBOL parser covers 10 construct types but does not handle the full COBOL-85 standard (which has approximately 250 reserved words). Notable unsupported constructs include \texttt{INSPECT}, \texttt{STRING}/\texttt{UNSTRING}, \texttt{SORT}, \texttt{ALTER}, \texttt{REDEFINES}, \texttt{OCCURS} (table handling), \texttt{COMP-3} (packed decimal), level-88 condition names, and nested programs. Several of these---particularly level-88 conditions, \texttt{REDEFINES}, and \texttt{OCCURS}---are ubiquitous in production COBOL. The current 10-construct scope covers a narrow slice of the language; extending the parser to handle these constructs is essential for practical applicability.

The correction loop is not evaluated in this paper (\S\ref{sec:feedback}).

\paragraph{The verifier as a check on our own generator.} The equivalence query judges our rule-based generator as well as LLM output: the six refutations of Table~\ref{tab:refuted} are defects of our generator that the query found. Because generator and IR share the parser, the query cannot find a defect in the parser itself (\S\ref{sec:noncircular}).

\paragraph{Baselines.} RQ5 compares \textsc{Ironproof}'s rule-based transpilation against LLM-only translation on the Z3-provable fragment, quantifying the value of the formal verification layer. However, we do not compare against commercial COBOL modernization tools (e.g., IBM watsonx Code Assistant for Z, AWS Mainframe Modernization), which would require access to proprietary tooling and production COBOL codebases. A comparison against these offerings on shared benchmarks remains necessary to fully establish \textsc{Ironproof}'s practical positioning.

\paragraph{Benchmark Scale.} The 2{,}345-program benchmark includes real, independently authored code from the GnuCOBOL compiler test suite and the NIST CCVS85 validation suite. However, 66.7\% of these programs yield no Z3-encodable output at all, a further 3.2\% fail inside our own pipeline, and only 14 of the 606 proofs quantify over an input that a proved output depends on. Production COBOL programs commonly span thousands of lines with complex inter-paragraph control flow, nested COPY books, and COMP-3 packed decimal. Extending the Z3-encodable fragment to cover loops, I/O effects, and string operations would substantially increase the provable proportion.

\subsection{Future Work}

Five directions would substantially extend \textsc{Ironproof}'s scope:

\paragraph{A Versioned Runtime Oracle.} The confrontation of \S\ref{sec:oracle} ran once, on an earlier pipeline, with a harness that was not versioned. Re-running it on the current pipeline, from a script in the artifact, comes before any other extension: it is the only measurement in this paper that checks the IR against a COBOL implementation.

\paragraph{Precise Fixed-Point Encoding.} The current Z3 encoding uses \texttt{Real} (infinite-precision rationals), which over-approximates COBOL's fixed-point semantics. Replacing this with Z3 \texttt{BitVec} constraints or bounded integers with explicit PIC-derived range constraints would enable detection of overflow and truncation bugs that the current encoding misses. This is the most critical improvement for the soundness of \textsc{Ironproof}'s formal guarantees.

\paragraph{Loop Invariant Synthesis.} For \texttt{PERFORM UNTIL} and \texttt{PERFORM VARYING} loops, an invariant-based approach could extend Z3 verification to iterative constructs. Given a loop \texttt{PERFORM UNTIL cond}, a candidate invariant $I$ would need to satisfy: (1) $I$ holds before the loop; (2) if $I \wedge \neg\text{cond}$ holds, then $I$ holds after one iteration; and (3) $I \wedge \text{cond}$ implies the desired postcondition. LLMs could be used to propose candidate invariants, with Z3 checking the three conditions, creating a nested LLM-Z3 loop.

\paragraph{Abstract Interpretation for I/O.} File I/O and SQL constructs could be approximated using abstract interpretation: rather than modeling the concrete file system or database, we could model the \emph{effect} of I/O operations on working storage variables using an over-approximate abstract domain, enabling partial verification of programs that mix computation with I/O.

\paragraph{Scaling to Full Programs.} Compositional verification---proving equivalence paragraph-by-paragraph and then composing the proofs---would allow \textsc{Ironproof} to handle larger programs. This requires modeling the calling conventions between paragraphs and ensuring that composed proofs remain sound.


\section{Related Work}\label{sec:related}

\paragraph{Legacy Modernization.} COBOL modernization has been studied for decades~\cite{sneed2010automated}. Traditional approaches rely on rule-based source-to-source transformation. IBM watsonx Code Assistant for Z~\cite{ibm2024watsonx} assists COBOL-to-Java transformation with an LLM, and AWS Mainframe Modernization~\cite{aws2023mainframe} offers tools to refactor mainframe applications. Pan et al.~\cite{pan2024cobol} study bugs introduced by LLMs when translating between C, C++, Go, Java and Python; COBOL is not among the languages they study. \textsc{Ironproof} focuses on machine-checkable equivalence proofs for the computational core of translated programs; we have not evaluated whether these products provide comparable checks.

\paragraph{LLM Code Translation.} TransCoder~\cite{roziere2020transcoder} demonstrated unsupervised programming language translation using back-translation on monolingual corpora. Chen et al.~\cite{chen2021evaluating} evaluate Codex, a model pretrained on code, on generating Python from natural-language descriptions. StarCoder~\cite{li2023starcoder} and other open models have further advanced the state of the art. Gao et al.~\cite{gao2023incontext} study what makes effective in-context demonstrations for code intelligence tasks. Jain et al.~\cite{jain2022jigsaw} combine LLMs with program synthesis for code generation. Pan et al.~\cite{pan2024cobol} categorize the bugs LLMs introduce when translating code between five languages. \textsc{Ironproof} takes the complementary route of checking each translation with a solver rather than characterizing errors after the fact.

\paragraph{Formal Verification of Translations.} Translation validation---verifying that a compiler or translator preserves semantics---originates with Pnueli et al.~\cite{pnueli1998translation} and has been realized in several practical systems. CompCert~\cite{leroy2009compcert} provides a fully verified C compiler whose correctness is proved in Coq. Alive2~\cite{lopes2021alive2} performs bounded translation validation for LLVM optimizations using SMT solving, and has found miscompilation bugs in LLVM. CertiCoq~\cite{anand2017certicoq} verifies compilation from Gallina (Coq's specification language) to C. These systems target compiler correctness within a single language family. \textsc{Ironproof} addresses a different problem---cross-language translation validation between COBOL and Python---which requires bridging a much larger semantic gap.

\paragraph{SMT-Based Program Equivalence.} Godlin and Strichman~\cite{godlin2009regression} formalize regression verification---proving that two versions of a program are semantically equivalent---using SMT solving. Unno et al.~\cite{unno2021constraint} develop constraint-based relational verification techniques for proving properties about pairs of programs. Lahiri et al.~\cite{lahiri2012symdiff} present SymDiff, a language-agnostic tool for checking program equivalence using verification conditions. \textsc{Ironproof} adapts these ideas to the specific context of COBOL-to-Python translation, with a bespoke Z3 encoding for COBOL's type system and construct semantics.

\paragraph{Positioning.} \textsc{Ironproof} sits at the intersection of three research threads that have largely evolved independently: legacy modernization (which needs automation), LLM code translation (which needs guarantees), and formal translation validation (which needs scalability). \textsc{Ironproof} combines rule-based generation, SMT-based checking and LLM-driven correction; we have not compared it with commercial modernization tools (\S\ref{sec:discussion}).


\section{Conclusion}\label{sec:conclusion}

We have presented \textsc{Ironproof}, a pipeline for COBOL-to-Python transpilation that combines rule-based Python emission with Z3-based formal equivalence verification and an LLM-driven feedback loop for error correction. \textsc{Ironproof} parses COBOL programs into a typed intermediate representation covering 10 constructs, generates Python with \texttt{Decimal}-typed arithmetic, encodes both programs as Z3 formulas over shared symbolic variables, and checks whether any input witnesses a semantic divergence.

On a benchmark of 2{,}345 COBOL programs---including 1{,}632 from the GnuCOBOL compiler test suite and 69 from the NIST CCVS85 validation suite---\textsc{Ironproof} proves equivalence on \textbf{606 of the 782 programs that enter the checking path (77.5\%)}, and on 153 of 291 (52.6\%) when the code is independently authored. Our own pipeline's 75 failures stay in the denominator. Two limits govern how far those proofs reach. 37 of the independent proofs cover a single execution of a program that reads input the encoder does not model, and only 14 proofs quantify over an input that a proved output depends on. And on two public business-application corpora, the encoder models no program end to end. The benchmark itself no longer refutes any translation of our generator, because the later hardening stops earlier the six programs it refuted on 2026-08-26; the obligation's capacity to fail is shown by those six and by 38 LLM translations. A baseline run in April 2026 shows that, on 94 independently authored programs, LLM-only translation failed on 57\%---Z3-confirmed semantic divergences together with translations our encoder could not verify. PIC-bounded overflow detection flags at least one out-of-range output in 22.1\% of the programs it can analyze---an upper bound, since the analysis quantifies over unconstrained inputs. Confronting the proved programs of the 2026-08-26 pipeline with GnuCOBOL 3.2.0 (\S\ref{sec:oracle}), 24 of the 49 we could execute end to end disagreed with the reference runtime on a variable inside the proof's own domain. An emission-fidelity proof is therefore a filter, not a substitute for confrontation with an implementation.

\textsc{Ironproof} does not claim to solve the full COBOL modernization problem. Its formal verification is limited to pure computational sections, its parser covers a narrow subset of COBOL-85, and comparison against commercial modernization tools remains necessary. Within its scope, it shows that SMT-based equivalence checking can provide machine-checkable proofs for cross-language translation validation, or concrete counterexamples where a translation diverges.

\begin{acks}
The author declares a financial interest in the system evaluated.
\end{acks}

\bibliographystyle{ACM-Reference-Format}
\bibliography{ironproof_cobol}


\begin{thebibliography}{20}


\ifx \showCODEN    \undefined \def \showCODEN     #1{\unskip}     \fi
\ifx \showDOI      \undefined \def \showDOI       #1{#1}\fi
\ifx \showISBNx    \undefined \def \showISBNx     #1{\unskip}     \fi
\ifx \showISBNxiii \undefined \def \showISBNxiii  #1{\unskip}     \fi
\ifx \showISSN     \undefined \def \showISSN      #1{\unskip}     \fi
\ifx \showLCCN     \undefined \def \showLCCN      #1{\unskip}     \fi
\ifx \shownote     \undefined \def \shownote      #1{#1}          \fi
\ifx \showarticletitle \undefined \def \showarticletitle #1{#1}   \fi
\ifx \showURL      \undefined \def \showURL       {\relax}        \fi
\providecommand\bibfield[2]{#2}
\providecommand\bibinfo[2]{#2}
\providecommand\natexlab[1]{#1}
\providecommand\showeprint[2][]{arXiv:#2}

\bibitem[{Amazon Web Services}(2026)]%
        {aws2023mainframe}
\bibfield{author}{\bibinfo{person}{{Amazon Web Services}}.}
  \bibinfo{year}{2026}\natexlab{}.
\newblock \bibinfo{title}{{AWS Mainframe Modernization}}.
\newblock \bibinfo{howpublished}{Product page}.
\newblock
\newblock
\shownote{\url{https://aws.amazon.com/mainframe-modernization/}, accessed
  2026-10-07}.


\bibitem[Anand et~al\mbox{.}(2017)]%
        {anand2017certicoq}
\bibfield{author}{\bibinfo{person}{Abhishek Anand}, \bibinfo{person}{Andrew
  Appel}, \bibinfo{person}{Greg Morrisett}, \bibinfo{person}{Zoe
  Paraskevopoulou}, \bibinfo{person}{Randy Pollack},
  \bibinfo{person}{Olivier~Savary Belanger}, \bibinfo{person}{Matthieu Sozeau},
  {and} \bibinfo{person}{Matthew Weaver}.} \bibinfo{year}{2017}\natexlab{}.
\newblock \showarticletitle{{CertiCoq: A Verified Compiler for Coq}}. In
  \bibinfo{booktitle}{\emph{CoqPL Workshop}}.
\newblock


\bibitem[Chen et~al\mbox{.}(2021)]%
        {chen2021evaluating}
\bibfield{author}{\bibinfo{person}{Mark Chen} {et~al\mbox{.}}}
  \bibinfo{year}{2021}\natexlab{}.
\newblock \bibinfo{title}{{Evaluating Large Language Models Trained on Code}}.
\newblock \bibinfo{howpublished}{arXiv:2107.03374}.
\newblock


\bibitem[de~Moura and Bj{\o}rner(2008)]%
        {demoura2008z3}
\bibfield{author}{\bibinfo{person}{Leonardo de Moura} {and}
  \bibinfo{person}{Nikolaj Bj{\o}rner}.} \bibinfo{year}{2008}\natexlab{}.
\newblock \showarticletitle{{Z3: An Efficient SMT Solver}}. In
  \bibinfo{booktitle}{\emph{Proceedings of the International Conference on
  Tools and Algorithms for the Construction and Analysis of Systems (TACAS)}}.
  \bibinfo{publisher}{Springer}, \bibinfo{pages}{337--340}.
\newblock


\bibitem[{Deloitte}(2024)]%
        {deloitte2024banking}
\bibfield{author}{\bibinfo{person}{{Deloitte}}.}
  \bibinfo{year}{2024}\natexlab{}.
\newblock \bibinfo{booktitle}{\emph{{EngineeringBeat: Navigating Mainframe
  Modernization Complexity---Overcoming Digital Transformation Challenges in
  Banking}}}.
\newblock \bibinfo{type}{{T}echnical {R}eport}. \bibinfo{institution}{Deloitte
  United States}.
\newblock
\newblock
\shownote{\url{https://www.deloitte.com/us/en/services/consulting/articles/mainframe-modernization-in-banking.html}}.


\bibitem[Gao et~al\mbox{.}(2023)]%
        {gao2023incontext}
\bibfield{author}{\bibinfo{person}{Shuzheng Gao}, \bibinfo{person}{Xin-Cheng
  Wen}, \bibinfo{person}{Cuiyun Gao}, \bibinfo{person}{Wenxuan Wang},
  \bibinfo{person}{Hongyu Zhang}, {and} \bibinfo{person}{Michael~R. Lyu}.}
  \bibinfo{year}{2023}\natexlab{}.
\newblock \showarticletitle{{What Makes Good In-Context Demonstrations for Code
  Intelligence Tasks with LLMs?}}. In \bibinfo{booktitle}{\emph{Proceedings of
  the International Conference on Automated Software Engineering (ASE)}}.
  \bibinfo{publisher}{IEEE/ACM}.
\newblock
\newblock
\shownote{arXiv:2304.07575}.


\bibitem[Godlin and Strichman(2009)]%
        {godlin2009regression}
\bibfield{author}{\bibinfo{person}{Benny Godlin} {and} \bibinfo{person}{Ofer
  Strichman}.} \bibinfo{year}{2009}\natexlab{}.
\newblock \showarticletitle{{Regression Verification}}. In
  \bibinfo{booktitle}{\emph{Proceedings of the 46th Design Automation
  Conference (DAC)}}. \bibinfo{publisher}{ACM}, \bibinfo{pages}{466--471}.
\newblock
\urldef\tempurl%
\url{https://doi.org/10.1145/1629911.1630034}
\showDOI{\tempurl}


\bibitem[{IBM}(2026)]%
        {ibm2024watsonx}
\bibfield{author}{\bibinfo{person}{{IBM}}.} \bibinfo{year}{2026}\natexlab{}.
\newblock \bibinfo{title}{{IBM watsonx Code Assistant for Z}}.
\newblock \bibinfo{howpublished}{Product page}.
\newblock
\newblock
\shownote{\url{https://www.ibm.com/products/watsonx-code-assistant-z}, accessed
  2026-10-07}.


\bibitem[Irrera(2017)]%
        {reuters2017cobol}
\bibfield{author}{\bibinfo{person}{Anna Irrera}.}
  \bibinfo{year}{2017}\natexlab{}.
\newblock \bibinfo{title}{{Banks scramble to fix old systems as IT `cowboys'
  ride into sunset}}.
\newblock \bibinfo{howpublished}{Reuters}.
\newblock


\bibitem[Jain et~al\mbox{.}(2022)]%
        {jain2022jigsaw}
\bibfield{author}{\bibinfo{person}{Naman Jain} {et~al\mbox{.}}}
  \bibinfo{year}{2022}\natexlab{}.
\newblock \showarticletitle{{Jigsaw: Large Language Models Meet Program
  Synthesis}}. In \bibinfo{booktitle}{\emph{Proceedings of the International
  Conference on Software Engineering (ICSE)}}. \bibinfo{publisher}{ACM/IEEE},
  \bibinfo{pages}{1219--1231}.
\newblock


\bibitem[Lahiri et~al\mbox{.}(2012)]%
        {lahiri2012symdiff}
\bibfield{author}{\bibinfo{person}{Shuvendu~K. Lahiri}, \bibinfo{person}{Chris
  Hawblitzel}, \bibinfo{person}{Ming Kawaguchi}, {and}
  \bibinfo{person}{Henrique Reb{\^e}lo}.} \bibinfo{year}{2012}\natexlab{}.
\newblock \showarticletitle{{SymDiff: A Language-Agnostic Semantic Diff Tool
  for Imperative Programs}}. In \bibinfo{booktitle}{\emph{Proceedings of the
  International Conference on Computer Aided Verification (CAV)}}.
  \bibinfo{publisher}{Springer}, \bibinfo{pages}{712--717}.
\newblock


\bibitem[Leroy(2009)]%
        {leroy2009compcert}
\bibfield{author}{\bibinfo{person}{Xavier Leroy}.}
  \bibinfo{year}{2009}\natexlab{}.
\newblock \showarticletitle{{A Formally Verified Compiler Back-end}}.
\newblock \bibinfo{journal}{\emph{Journal of Automated Reasoning}}
  \bibinfo{volume}{43}, \bibinfo{number}{4} (\bibinfo{year}{2009}),
  \bibinfo{pages}{363--446}.
\newblock


\bibitem[Li et~al\mbox{.}(2023)]%
        {li2023starcoder}
\bibfield{author}{\bibinfo{person}{Raymond Li} {et~al\mbox{.}}}
  \bibinfo{year}{2023}\natexlab{}.
\newblock \showarticletitle{{StarCoder: May the Source Be with You!}}
\newblock \bibinfo{journal}{\emph{Transactions on Machine Learning Research}}
  (\bibinfo{year}{2023}).
\newblock


\bibitem[Lopes et~al\mbox{.}(2021)]%
        {lopes2021alive2}
\bibfield{author}{\bibinfo{person}{Nuno~P. Lopes}, \bibinfo{person}{Juneyoung
  Lee}, \bibinfo{person}{Chung-Kil Hur}, \bibinfo{person}{Zhengyang Liu}, {and}
  \bibinfo{person}{John Regehr}.} \bibinfo{year}{2021}\natexlab{}.
\newblock \showarticletitle{{Alive2: Bounded Translation Validation for LLVM}}.
  In \bibinfo{booktitle}{\emph{Proceedings of the ACM SIGPLAN Conference on
  Programming Language Design and Implementation (PLDI)}}.
  \bibinfo{publisher}{ACM}, \bibinfo{pages}{65--79}.
\newblock


\bibitem[{National Institute of Standards and Technology}(1986)]%
        {nist1985cobol}
\bibfield{author}{\bibinfo{person}{{National Institute of Standards and
  Technology}}.} \bibinfo{year}{1986}\natexlab{}.
\newblock \bibinfo{booktitle}{\emph{{COBOL Compiler Validation System
  (CCVS85)}}}.
\newblock \bibinfo{type}{{T}echnical {R}eport}. \bibinfo{institution}{NIST}.
\newblock
\newblock
\shownote{User guide; test suite version 4.2, 1993}.


\bibitem[Pan et~al\mbox{.}(2024)]%
        {pan2024cobol}
\bibfield{author}{\bibinfo{person}{Rangeet Pan}, \bibinfo{person}{Ali~Reza
  Ibrahimzada}, \bibinfo{person}{Rahul Krishna}, \bibinfo{person}{Divya
  Sankar}, \bibinfo{person}{Lambert~Pouguem Wassi}, \bibinfo{person}{Michele
  Merler}, \bibinfo{person}{Boris Sobolev}, \bibinfo{person}{Raju Pavuluri},
  \bibinfo{person}{Saurabh Sinha}, {and} \bibinfo{person}{Reyhaneh
  Jabbarvand}.} \bibinfo{year}{2024}\natexlab{}.
\newblock \showarticletitle{{Lost in Translation: A Study of Bugs Introduced by
  Large Language Models while Translating Code}}. In
  \bibinfo{booktitle}{\emph{Proceedings of the IEEE/ACM 46th International
  Conference on Software Engineering (ICSE)}}. \bibinfo{publisher}{ACM/IEEE},
  \bibinfo{pages}{1--13}.
\newblock
\urldef\tempurl%
\url{https://doi.org/10.1145/3597503.3639226}
\showDOI{\tempurl}


\bibitem[Pnueli et~al\mbox{.}(1998)]%
        {pnueli1998translation}
\bibfield{author}{\bibinfo{person}{Amir Pnueli}, \bibinfo{person}{Michael
  Siegel}, {and} \bibinfo{person}{Eli Singerman}.}
  \bibinfo{year}{1998}\natexlab{}.
\newblock \showarticletitle{Translation validation}. In
  \bibinfo{booktitle}{\emph{Tools and Algorithms for the Construction and
  Analysis of Systems (TACAS)}} \emph{(\bibinfo{series}{LNCS},
  Vol.~\bibinfo{volume}{1384})}. \bibinfo{publisher}{Springer},
  \bibinfo{pages}{151--166}.
\newblock


\bibitem[Rozi{\`e}re et~al\mbox{.}(2020)]%
        {roziere2020transcoder}
\bibfield{author}{\bibinfo{person}{Baptiste Rozi{\`e}re},
  \bibinfo{person}{Marie-Anne Lachaux}, \bibinfo{person}{Lowik Chanussot},
  {and} \bibinfo{person}{Guillaume Lample}.} \bibinfo{year}{2020}\natexlab{}.
\newblock \showarticletitle{{Unsupervised Translation of Programming
  Languages}}. In \bibinfo{booktitle}{\emph{Advances in Neural Information
  Processing Systems (NeurIPS)}}. \bibinfo{pages}{20601--20611}.
\newblock


\bibitem[Sneed(2010)]%
        {sneed2010automated}
\bibfield{author}{\bibinfo{person}{Harry~M. Sneed}.}
  \bibinfo{year}{2010}\natexlab{}.
\newblock \showarticletitle{{Migrating from COBOL to Java}}. In
  \bibinfo{booktitle}{\emph{IEEE International Conference on Software
  Maintenance (ICSM)}}. \bibinfo{publisher}{IEEE}, \bibinfo{pages}{1--7}.
\newblock
\urldef\tempurl%
\url{https://doi.org/10.1109/ICSM.2010.5609583}
\showDOI{\tempurl}


\bibitem[Unno et~al\mbox{.}(2021)]%
        {unno2021constraint}
\bibfield{author}{\bibinfo{person}{Hiroshi Unno}, \bibinfo{person}{Tachio
  Terauchi}, {and} \bibinfo{person}{Eric Koskinen}.}
  \bibinfo{year}{2021}\natexlab{}.
\newblock \showarticletitle{{Constraint-based Relational Verification}}. In
  \bibinfo{booktitle}{\emph{Proceedings of the International Conference on
  Computer Aided Verification (CAV)}}. \bibinfo{publisher}{Springer},
  \bibinfo{pages}{742--766}.
\newblock


\end{thebibliography}

\appendix

\section{LLM Prompts Used in the Feedback Loop}\label{app:prompts}

For reproducibility, this appendix presents the translation prompt and the correction prompt of \textsc{Ironproof} (\S\ref{sec:feedback}). The only LLM results in this paper, the baseline of \S\ref{sec:rq6}, were produced with \texttt{claude-opus-4-7}, recorded in the raw result files under \texttt{benchmark/results/}. We name it rather than describing it generically: a translation failure rate is a property of a specific model at a specific date, and is not interpretable without both.

\subsection{System Prompt for Initial Translation}

The initial translation prompt instructs the LLM to translate COBOL to Python while preserving exact semantics. It is reproduced below (verbatim except that the arrow is typeset as \texttt{->}), and is the prompt of the baseline of \S\ref{sec:rq6}. The tool and the baseline build it from the same template (\texttt{ironproof\_core.py}, \texttt{translate\_with\_\allowbreak claude} and \texttt{benchmark/\allowbreak run\_baseline.py}, \texttt{translate\_\allowbreak llm}; the two are character-identical):

\begin{lstlisting}[style=z3style,caption={System prompt for COBOL-to-Python translation, verbatim from the implementation.},label={lst:system-prompt},float=tbp]
Expert COBOL->Python migration. Translate exactly. Rules:
- Single function named after PROGRAM-ID (snake_case)
- All variables use Decimal for numeric types
- Preserve exact conditions (<=, <, etc.)
- Use same variable names (lowercase)
- No comments, no imports

COBOL:
```
[COBOL SOURCE INSERTED HERE]
```
Python only, no explanation.
\end{lstlisting}

\subsection{Correction Prompt with Z3 Counterexample}

When Z3 returns SAT, \textsc{Ironproof} extracts the counterexample and builds the following correction prompt:

\begin{lstlisting}[style=z3style,caption={Correction prompt template, from \texttt{translate\_with\_feedback\_loop} in \texttt{ironproof\_core.py}; bracketed parts are filled at run time.},label={lst:correction-prompt},float=tbp]
Your COBOL->Python translation has a semantic bug.

The Z3 SMT solver found a counterexample proving the
COBOL and Python produce different outputs:

[for each refuted output:]
Variable <NAME>: <solver details>
  Counterexample inputs: <input assignment>

Original COBOL:
```
[COBOL SOURCE]
```

Your buggy Python:
```python
[CURRENT PYTHON]
```

Fix the Python translation so it is semantically
equivalent to the COBOL. Pay close attention to:
- Boundary conditions (<=, <, >=, >)
- Operator precedence
- Decimal precision
- Variable initialization

Return ONLY the corrected Python code, no explanation.
\end{lstlisting}

\end{document}